\documentclass[11pt, a4paper]{article} 

\usepackage[german,english]{babel} 
\usepackage{booktabs} 
\usepackage{comment} 
\usepackage[utf8]{inputenc} 
\usepackage[T1]{fontenc} 

\usepackage{amsmath,amsfonts,amsthm} 
\usepackage{tikz} 
\usepackage{tikz-cd}
\usetikzlibrary{positioning,arrows} 
\usetikzlibrary{decorations.pathreplacing} 
\usepackage{tikzsymbols} 
\usepackage{blindtext} 

\usepackage{geometry} 

\usepackage{setspace} 
\usepackage[T1]{fontenc} 
\usepackage[utf8]{inputenc} 

\usepackage{XCharter} 

\usepackage{hyperref}  

\usepackage{fancyhdr} 
\usepackage[square,sort,comma,numbers]{natbib}
\usepackage{har2nat} 

\usepackage{authblk}

\title{Exploiting the IRT-THESEUS capability  to observe  lensed Quasars
} 

\author[1]{L. Hamolli\footnote{Contact author: Lindita Hamolli: lindita.hamolli@fshn.edu.al}}
\author[1]{M. Hafizi}
\author[2,3]{F. De  Paolis}
\author[2,3]{A. A. Nucita}

\affil[1]{Department of Physics, University of Tirana, Albania}
\affil[2]{Department of Mathematics and Physics "Ennio De Giorgi", University of Salento, I-73100 Lecce, Italy}
\affil[3]{INFN, Sezione di Lecce, CP 193, I-73100 Lecce, Italy}

\date{\small \today} %

\begin{document}


\maketitle 

\begin{center} 
\end{center}

\begin{abstract}
{THESEUS is an ESA space based project, aiming to explore the early universe by unveiling a complete census of Gamma-Ray Burst (GRB) population in the first billion years. This goal is expected to be achieved by combined observations of its three instruments: the Soft X-ray Imager (SXI), the X and Gamma Imaging Spectrometer (XGIS) and the InfraRed Telescope (IRT). In particular, the IRT instrument will help to identify, localise and study the afterglow of the GRBs detected by SXI and XGIS, and about $40\%$ of its time  will be devoted to an all-sky photometric survey, which will certainly detect a relevant number of extragalactic sources, including Quasars. In this paper, we focus on the capability of IRT-THESEUS Telescope to observe Quasars and, in particular, those objects lensed by foreground galaxies. In our analysis, we consider the Quasar Luminosity Function  (QLF) in the infrared band based obtained by the Spitzer Space Telescope  imaging survey. Furthermore, 
by using the mass-luminosity distribution function of galaxies and the galaxy/Quasar redshift distributions, we preformed Monte Carlo simulations  to estimate the number of lensed Quasars.
We predict that up to  $2.14 \times 10^5$ Quasars can be observed during IRT-Theseus sky survey, and about $140$ of them could be lensed by foreground galaxies. Detailed studies of these events would provide a powerful probe of the physical properties of Quasars and the mass distribution models of the galaxies.}


\end{abstract}

\setcounter{page}{1} 

\section{Introduction} 
THESEUS is an ESA space based project. Its specific scientific objective is to explore the physical conditions of the Early Universe by unveiling a complete census of the GRB population in the first billion years since the Big Bang and it can also perform an unprecedented deep monitoring of the X-ray transient Universe. THESEUS will observe several hundreds of GRBs  per year \cite{gotz2018infra}.

During the time of waiting for any GRB alert, THESEUS will be able to create a rich database about objects at high $z$, due to its three instruments on board, dedicated to $\gamma$, $X$ and $IR$ bands \cite{amati2018THESEUS}.

In this paper, we aim to  consider in particular the information about Quasars gathered by the instruments on-board THESEUS. Quasars are very energetic and distant galaxies with an active galactic nucleus (AGN). Being  distant and with extremely high luminosity, Quasars are  considered to be important instruments for discovering the content and the history of Universe. In this respect, one of the source of information is related to Quasars lensed by foreground galaxies.

Quasars were discovered in the 1960s and  Barnothy  \cite{barnothy1965quasars} was the first to connect them with the gravitational lens effect. However,  Quasar lensing  received a real boost only in 1979, when Walsh et al. \cite{walsh19790957+} discovered  the first double Quasar.


Concerning the gravitational lensing, General Relativity predicts the value of the deflection angle $\alpha$ of a light ray by a spherically symmetric mass M. If the impact parameter $\xi$ (the smaller distance of the light ray with respect to the deflecting body) is much larger than the Schwarzschild radius of the lens mass ($\xi \gg R_S \equiv \frac{2GM}{c^2}$), the deflection angle is given by
\begin{equation}
 \alpha=\frac{4GM}{c^2 \xi}.  
\end{equation}
This is exactly twice the value obtained in Newtonian gravity. The deflection angle of an ensemble of mass points is the sum of the deflections due to individual mass components, which in the case of the  geometrical thin lens (the extent of the deflecting mass along the line of sight is much smaller than  both its distances to the observer and the source) is given by \cite{schneider2006gravitational}:
\begin{equation}
 \alpha(\vec{\xi})=\frac{4G}{c^2}\int d^2\vec{\xi'}\Sigma(\vec\xi')\frac{\vec{\xi}-\vec{\xi'}}{|\vec{\xi}-\vec{\xi'}|^2} .
\end{equation}
Here,  $\Sigma(\vec{\xi})$ is the mass density projected onto a plane perpendicular to the incoming light ray from the source and arriving to the observer (line of sight). 

From here, the so-called lens equation is found to be \cite{de2016scales}

\begin{equation}
 \theta- \beta=\theta_E^2/ \theta,
 \label{lenseq}
 \end{equation}
where $\beta$ is the angular position of the source, $\theta$ are the angular positions of the source images and  $\theta_E$ is the Einstein ring radius, the angular radius of the image when the lens and the
source are perfectly aligned. It is given by 
\begin{equation}
\theta _E=\sqrt{\frac{4GM(R_E)}{c^2}\frac{D_{LS}}{D_S D_L}},
\label{teta}
\end{equation}
where {$M(R_E)$ is the lens galaxy mass inside the Einstein radius } and $D_S$, $D_L$, $D_{LS}$ are the angular diameter distances observer-source, observer-lens and lens-source, respectively.

In cosmology, the angular diameter distance of a source at redshift $z_2$ seen by an observer at redshift $z_1$ is given by \cite{krawczynski2019effect},
\begin{equation}
  D(z_1,z_2)=\frac{c}{H(z_1)}\frac{1+z_1}{1+z_2}\int_{z_1}^{z_2}\frac{dz'}{E(z')} 
   \label{}
\end{equation}
with 
\begin{equation}
  E(z)=\sqrt{\Omega_m(1+z)^3+\Omega_k(1+z)^2 +\Omega_{\Lambda}}~~~and ~~~ H(z)=H_0E(z). 
  \label{E(z)}
\end{equation}
Here, $H_0$ is Hubble constant and  $\Omega_m$,  $\Omega_k$, $\Omega_{\Lambda}$ are the dimensionless density parameters for the matter (i.e. the sum of the cold dark matter and baryonic matter, the space-curvature, and  the dark energy, respectively).

There are three different scales in gravitational lensing: strong gravitational lensing (when the lens images are separated by more than a few tenths of arcsecs and can be observed as distinct images);
weak lensing (when the involved light deflection angles are much smaller and galaxy distortion can be observed only  statistically over a large number of galaxies); and microlensing (in the case of star-on-star lensing, where the angular separation between images is of the order of $\mu$arcsec)\footnote{See, e.g. \cite{de2016scales} for a review on the subject.}. 

 Gravitationally lensed Quasars by foreground galaxies belong to the strong lensing scale. They come in a variety of classes: double, triple and quadruple systems, { observed up to now by several space- and ground-based telescopes (Hubble Space Telescope (HST) \cite{keeton2000host}, Sloan Digital Sky Survey (SDSS) \cite{inada2012sloan}, Chandra \cite{walton2015broad}, Spitzer \cite{williams2016spitzer}  etc)}. Their study provides a lot of precious information about the lens (foreground galaxies), the source (Quasars) and allows a better understanding of galactic and Quasar populations in the Universe, their distribution in distance (or redshift) and mass. 
 
 Recently, these studies lead even to the discovery of exoplanets in faraway galaxies \cite{dai2018probing}.
Indeed, strongly lensed Quasars may also be affected by microlensing \footnote{Quasars might also sometime be microlensed by e.g. a star in the Quasar hosting galaxy as possibly happened recently in the event towards J1249+3449. These kind of events would give rise to an X-ray flare, analogous to the optical flare associated to that event \cite{graham,de2020quasar}.}, since light bundles from the lensed Quasar may go through the lensing galaxy and/or its galactic halo. Galaxies consist at least partly of stars, and galactic halos consist possibly of compact objects in the mass range $[10^{-6}, 10^3]M_{\odot}$, as well. Each of these stars (or other compact objects, like black holes, brown dwarfs, or planets) may act as a “compact lens” or “microlens” and produces 
new images of the source. But, because the image splitting is proportional to the square root of the lens mass, these microimages are only of the order a  $\mu$arcsec  apart and can not be directly resolved by any of the  up to date available techniques. Instead, a characteristic light-curve is observed.
Recent studies focused on microlensing examination show that sometimes the variation in the Quasar light could be a source of information for the presence of compact objects in the lensing galaxy or in the galaxy hosting the Quasar \cite{de2020quasar}.

{In this work we aim to examine the capacities of IRT-THESEUS to observe Quasars and in particular Quasars lensed by foreground galaxies. The structure of this paper is as follows: in Section 2 we predict and the number of Quasar observed in a IR band close to the band of IRT-THESEUS. In Section 3 we describe the algorithm which we established and used to estimate the number of lensed Quasars by foreground galaxies, observed by any telescope and explain how we find the estimated number of lensed Quasars, observed by IRT-THESEUS. In this section we show the test of this algorithm on the data of the SDSS and of the Chandra X-ray telescope. Our conclusions are drawn in Section 4.  \\
In our calculations we use the following cosmological parameters: $\Omega_m=0.30$, $\Omega_k=0$, $\Omega_\Lambda = 0.70$ and $H_0=70$ km s$^{-1}$ Mpc$^{-1}$.}

\section{Quasars as expected to be observed by IRT-THESEUS} 

In this section, we examine the  possibility of the IR instrument on board of THESEUS to observe Quasars. IRT-THESEUS is designed to be a 0.7 m class telescope with $15\times 15$ arcmin FOV, with both imaging and spectroscopy capabilities in the $0.7-1.8$ $\mu m$ band \cite{gotz2018infra}, designed to operate in three observation modes: the photometric mode,  the low-resolution (LR) mode and  the high-resolution (HR) mode. When a gamma-ray burst is identified, the IRT instrument will execute the following observational sequence: it will observe in imaging mode  within $1$ arcsec of the GRB error box and by frames in ZJH photometric bands so that the source colour can be determined; then  IRT will enter the spectroscopy mode to define the redshift value and the position of the source.  In the case of a bright source, it can go back to imaging mode (H-band) for at least another 1800s. When the source is faint, IRT takes data in imaging mode for a 3600s time interval to establish the GRB photometric light curve \footnote{https://www.isdc.unige.ch/theseus/mission-payload-and-profile.html}. The rest of the time, IRT-THESEUS will be able to perform an all-sky photometric survey and detect a number of extragalactic sources, including Quasars.

Here we estimate the number of Quasars expected to be detected and recorded on the  IRT-THESEUS database, based on the instrument's sensitivity.

\subsection{Number of Quasars observed in IR Band}
Quasars are luminous in almost all accessible bands. The Quasar luminosity function (QLF), the comoving number density of Quasars as a function of luminosity, is perhaps the most important observational signature of Quasar populations. Lacy et al. \cite{lacy2015spitzer} based on the Spitzer Space Telescope imaging surveys derived the luminosity functions of active galactic nuclei (AGNs) in IR band as a double power-law function of the following form:
\begin{equation}
 \phi (L,z)=\frac{d\phi}{dlogL}=\frac{\phi^{\ast}}{\left[\left( \frac{L}{L^{\ast}}\right)^{\gamma_1}+\left( \frac{L}{L^{\ast}}\right)^{\gamma_2}\right]},   
\end{equation}
where $\phi$ is the comoving space density of the AGN,   $\phi^{\ast}$ is the
characteristic space density, both in units of comoving $Mpc^{-3}$, $L$ is the rest-frame luminosity at $5 ~\mu m$, and $L^{\ast}$ is the break luminosity at $5~\mu m$, both in units of $erg~s^{-1} Hz^{-1}$. The evolution in $L^{\ast}$  is  a cubic  expression  (e.g.,\cite{hopkins2007observational} )
\begin{equation}
log_{10}L^{\ast}(z)=log_{10}L^{\ast}_0+k_1\epsilon+k_2\epsilon^2+k_3\epsilon^3,
\end{equation}
where $\epsilon=log_{10}((z+1)/(z+z_{ref}))$, $L^{\ast}_0$ is a free parameter in
the fit, corresponding to the break luminosity at redshift zero, $\gamma_1$
is the faint-end slope, $\gamma_2$ is the bright-end slope and $z_{ref}$ is fixed at $2.5$.
The coefficients $k_1$, $k_2$, $k_3$, $\gamma_1$, $\gamma_2$, $\phi^{\ast}$ and  $L^{\ast}_0$  are given in the Tab. 1 of Lacy et al. \cite{lacy2015spitzer}.

In order to determine the number of Quasars in the redshift range $z_1-z_2$, we calculate the integral
\begin{equation}
N=\int_{z_1}^{z_2} \int_{L_{min}}^{L_{max}} \phi(L,z) dL dV_c  
\label{totnum}
\end{equation}
where $dV_c=D_H\frac{(1+z)^2 D_A^2}{E(z)}d\Omega dz$ is the comoving volume element in the solid angle $d\Omega $ and redshift interval $dz$, $D_H=\frac{c}{H_0}$ is the Hubble distance,  $D_A$ is the angular diameter distance at redshift $z$ and $E(z)$ is defined through the equation (\ref{E(z)}).

\subsubsection{A test on Spitzer Space Telescope }

Spitzer Space Telescope observed the Universe in several bands in mid-infrared wavelengths. Based on its capability, we consider the case of the band $3.6~ \mu m$  (the closest to the IRT-THESEUS band of observation), with apparent magnitude  18  \cite{williams2016spitzer}  and  calculate the number given by equation (\ref{totnum}) to be tested. For this magnitude, we find the minimum luminosity to be, 
 $L_{min}=L_{\odot}10^{\left(-17.88-log\frac{23.04x10^{-24}}{D_L^2}\right)}$,  where $L_{\odot}=3.83\times 10^{33}~~erg~s^{-1}$ is the solar luminosity and $D_L$ is the luminosity distance of the source. Considering $L_{max}=10^{47}erg \,s^{-1}$  in  equation  (\ref{totnum}),  we estimate the number of Quasars observed by Spitzer in the whole sky survey to be  $N\simeq 2.55\times 10^4$. At the same time, equation (\ref{totnum}) limits Spitzer observation of Quasars up to $z=2.58$. To check the number $N\simeq 2.55\times 10^4$, we find in Lacy and Sajina  \cite{lacy2020active} that over $20,000$ AGN candidates exist in the Spitzer archive, which is close to the theoretical value found above.  


\subsubsection{Results on IRT-THESEUS}

In the case of IRT-THESEUS, we are based on the instrument description \cite{gotz2018infra}: the AB mag (300s) is designed to be 20.6 for H band. Using the Vega - AB Magnitude Conversion, we convert this value to the value 19.21 of the apparent magnitude. In such conditions, we find $L_{min}=L_{\odot}10^{\left(-18.396-log\frac{23.04x10^{-24}}{D_L^2}\right)}$,  where $D_L$ is the luminosity distance of the source. Considering  $L_{max}=10^{47}erg\, s^{-1}$ in equation (\ref{totnum}), we calculate the number of Quasars to be observed by the IRT-THESEUS (in the overall field of view): $N\simeq 3.35\times 10^5$. At the same time, equation (\ref{totnum}) leads to a limit $z=4.23$ for the redshift 
of the Quasars observed by this instrument.\\
With THESEUS Field of Regards (FoR) $64\%$ (the fraction of the sky to be monitored) \cite{amati2018THESEUS}, the number of Quasars to be observed by THESEUS reduces to  $N\simeq 2.14\times 10^5$.
We note that this high number provides an opportunity to detect also lensing effects on Quasars.
For this reason, we devoted a thorough analysis in order to estimate the number of lensed Quasars by foreground galaxies in the IRT-THESEUS database, which is shown in the next section.


\section{Quasar lensing with IRT-THESEUS}
Gravitational lensing  of Quasars is a powerful tool for different kinds of investigations in astrophysics and cosmology.  It occurs whenever the source (Quasar), the lens (foreground galaxy)
and the observer  are so well aligned that the observer-source direction lies inside the so-called Einstein radius of the lens \cite{schneider2006gravitational}.  About one out of $500$ Quasars has multiple images as a result of lensing by  a foreground galaxy \cite{schmidt2010quasar}. Below we describe an algorithm to estimate the number of these lensed Quasars to be observed by IRT-THESEUS telescope.

\subsection{The algorithm generating Quasars and galaxies}

We assume that a strong lensing event is detectable by the telescope if the image separation, which is of the order of the Einstein angle, is higher than the accuracy of the telescope. For investigating the possibility to observe such events, we establish a method based on Monte Carlo numerical simulations. For each generated event we extract, via the procedure detailed below, all needed parameters. In particular: \\
1 - We generate the redshift of the Quasar, following the Quasar distribution obtained by Schneider et al.  \cite{schneider2005sloan} (see section 3.1.1 below).\\
2 - We generate the redshift of the galaxy following the distribution obtained by  Appenzeller et al. \cite{appenzeller2004exploring} (see section 3.1.2 below), provided that the galactic redshift is lower than the Quasar redshift. The number of galaxies fulfilling this condition for each Quasar is found by the cumulative number of galaxies.\\
3 - For each generated galaxy we extract its mass by the stellar mass function \cite{davidzon2017cosmos2015} (see section 3.1.3 below) and based on the relation between the stellar mass of the galaxy and stellar velocity distribution obtained by Zahid et al. \cite{zahid2016scaling} (see section 3.1.4 below) we find its velocity dispersion.\\
4 - Based on the Singular Isothermal Sphere model for the galaxies (see section 3.1.5 below), and making use of the galaxy and Quasar redshifts generated above, we define the Einstein angle $\theta_E$ for the couple Quasar/galaxy \cite{schmidt2010quasar}.\\
Is the galaxy inside the Einstein angle? For an isolated galaxy, its probability to be inside this Einstein angle has the tiny value of $\theta_E^2/4$. But, considering 200 billions of galaxies in the observable Universe, this probability raises to $\simeq 10^{11}\theta_E^2/2$.  \\
We take into consideration a sample of $7000$ galaxies, so we repeat the procedure of galaxy's parameter generation $7000$ times \cite{appenzeller2004exploring}. In such conditions, the probability for each synthetic galaxy to be inside the Einstein angle turns out to be $10^8\theta_E^2/14$.\\
The code extracts a uniformly distributed number $n$ in the interval $(0,1)$. It keeps the case when $n~<10^8~\theta_E^2/14$ (aligned couple Quasar/galaxy), otherwise it rejects it (nonaligned couple Quasar/galaxy). \\
This procedure is repeated as many times as the number of galaxies  with redshift lower than the Quasar redshift.\\
In the case of any aligned couple Quasar-galaxy, the  Einstein angle is then compared with the angular accuracy of the instrument.\\
The procedure must be repeated as many times as the number of Quasars observed by the instrument.

{
Below we explain the  distributions mentioned above and find the expression of the Einstein angular radius when  consider a singular isothermal sphere (SIS) model for the galaxy.
After, we follow with estimating the number of lensed Quasars in the case of three instruments: SDSS, Chandra and IRT-THESEUS. }

\subsubsection{Quasar redshift distribution}

In \cite{schneider2005sloan}, Schneider et al. found a Quasar redshift distribution based on the SDSS Data Catalog (Third Data Release). This catalog contains $46,420$ Quasars with spectroscopic redshifts from $0.08$ to $5.41$. The area covered by the catalog is $4188 ~deg^2$. A clear majority of Quasars have redshifts below $z=2$, the median redshift is $1.47$, (the mode is $1.85$). There are $520$ Quasars at redshifts greater than four, of which $17$ are at redshifts greater than five. The catalog contains $69$ Quasars with redshifts below $0.15$. 

\subsubsection{Galaxy redshift distribution}
Appenzeller et al. \cite{appenzeller2004exploring} found a photometric redshift distribution of $7,000$ galaxies. They analyzed  distant galaxies observed near the South Galactic Pole by the FORS Deep Field (ESO VLT), in a sky area of  about $7x7$ arcmin$^2$. This distribution was also confirmed by Davison et al. \cite{davidzon2017cosmos2015}. 

\subsubsection{Galaxy mass distribution}
Davidzon et. al \cite{davidzon2017cosmos2015}, based on the COSMOS2015 catalogue, provided a comprehensive view of the galaxy stellar mass assembly between $z = 0.1$ and $6$. They fitted measurements  with double (up to $z=3$) and a single (beyond that bin) Schechter function, given by
\begin{equation}
\Phi(M)dM=
\left[\Phi_1^*\left(\frac{M}{M_*}\right)^{\alpha_1}+\Phi_2^*\left(\frac{M}{M_*}\right)^{\alpha_2}\right]\exp{\left(-\frac{M}{M_*}\right)}\frac{dM}{M_*}.
\end{equation}
Schechter parameters of the COSMOS2015 galaxy stellar mass function (SMF) are found in the Table 1 of the \cite{davidzon2017cosmos2015}.

\subsubsection{Galactic velocity dispersion}

Based on an analysis of SDSS and SHELS data, Zahid et al. \cite{zahid2016scaling} proposed a relation between the central stellar velocity dispersion $\sigma$ and the stellar mass $M$ of  the galaxies, which  is a power law with a break point
\begin{equation}
\left\{\begin{array}{cc}
  \sigma(M)= & \sigma_b\left(\frac{M}{M_b}\right)^{\alpha_1}~~~~for~~~M\leq M_b
  \\&\\
  \sigma(M)= & \sigma_b\left(\frac{M}{M_b}\right)^{\alpha_2}~~~~for~~~M> M_b.
\end{array}\right.
\label{Lorentz}
\end{equation}
The central stellar velocity dispersion is a measure of the stellar kinematics and is directly related to the gravitational potential of the system.
The fit parameters are: $log(M_b/M_{\odot})=10.26$, $log(\sigma_b)=2.073$, $\alpha_1=0.403$ and $\alpha_2=0.293$. The two indices $\alpha_1$ and $\alpha_2$ define the power law below and above the break point, respectively.  

\subsubsection{The Singular Isothermal Distribution}
For the lensing properties of galaxies we assume a singular isothermal sphere (SIS) distribution.  This spherical mass distribution yields flat rotation curves, such as observed for spiral galaxies. Their density distribution is described by \cite{schneider2006gravitational}
\begin{equation}
\rho(r)=\frac{\sigma_{SIS}^2}{2\pi G r^2}.
\label{distributin}
\end{equation}
Physically this model corresponds to a distribution of self-gravitating particles, where the velocity distribution at all radii is a Maxwellian with
one-dimensional velocity dispersion $\sigma_{SIS}$.   
The mass distribution (\ref{distributin}) has two limitations: the central density diverges as $\rho \propto r^{-2}$ (hence the name ‘singular’), and the total mass of this distribution diverges as $r \rightarrow \infty$. But, these do not dismiss our assumption. First, although real galaxies have nonzero core radii, they are generally too small (compared to the critical impact parameter) to make much difference.  Second, since we are restricting our attention to lenses which produce multiple images, our results will be very insensitive to the mass distribution outside of the critical impact parameter and  the calculation does not depend on the existence or nature of extended massive halos, but only on the well determined (by rotation curves) mass distribution in the inner regions of galaxies.

 By projecting the three-dimensional density along the line of sight, we obtain the corresponding surface density
\begin{equation}
\Sigma (\xi)=2\frac{\sigma_{SIS}^2}{2\pi G }\int_0^\infty \frac{dz}{\xi^2+z^2} =\frac{\sigma_{SIS}^2}{2 G  \xi}.   
\end{equation}
The mass of the lens inside a radius $R$ from the galactic center in the galactic plane is given by 
\begin{equation}
M(R)=\int_0^R 2\pi \xi \Sigma(\xi)d \xi= \frac{\pi}{G}\sigma_{SIS}^2R
\end{equation}
and, by replacing in the lens equation  ($\ref{teta}$) we have
\begin{equation}
 \theta_E=4\pi \frac{\sigma_{SIS}^2}{c^2} \frac{D_{LS}}{D_S} \label{theta} 
\end{equation}

Since the central stellar velocity dispersion $(\sigma)$ in equation (\ref{Lorentz}) is a measure of the stellar kinematics related to the gravitational potential of the system, we have considered it the same to $\sigma_{SIS}$ in equation (\ref{theta}). 

Anyway, we have to remark that Ofek et al. \cite{ofek2003redshift} predicted a $\sigma_{SIS}$ not exactly equal to the observed stellar velocity dispersion $(\sigma)$,  for the reason of the dark matter halos, which show to be dynamically hotter than luminous stars. This conclusion is based on X-ray observations. Furthermore, they propose $\sigma_{SIS}=f_e\sigma$,  where $(0.8)^{1/2}<f_e<(1.2)^{1/2}$.

\subsection{Test on SDSS and Chandra}

\begin{figure}[]
\includegraphics[width=10.5 cm]{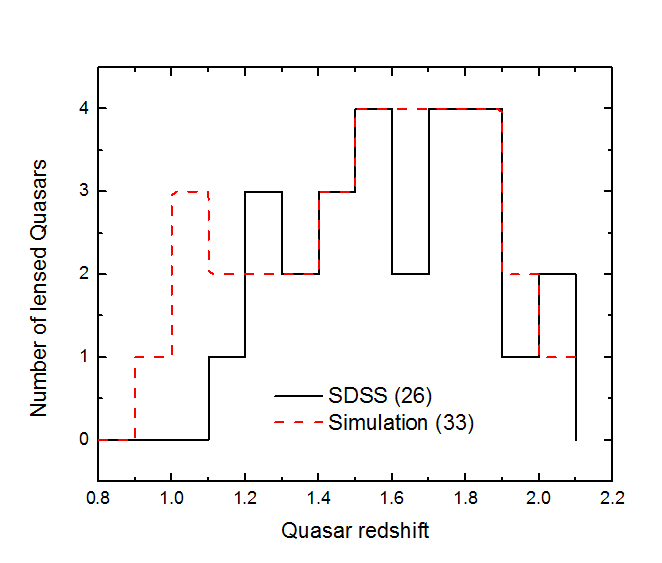}
\caption{Histogram of the redshift distributions for 26 lensed Quasars observed by the SDSS (continuous line) and for 33 synthetic lensed Quasars (dashed line).}
\label{SDSS}
\end{figure}

{We made a test by comparing our results with the observations conducted by the SDSS and Chandra X-ray telescope.  

In the case of SDSS, we consider a statistical sample of lensed Quasars\cite{inada2012sloan} consisting of 26 lensed Quasars brighter than $i=19.1$, with redshift in the range $0.6 < z < 2.2$,  selected from $50,836$ spectroscopically confirmed Quasars in the SDSS Data. The image separation range is restricted  to $1< \theta < 20~$arcsec. \\
We consider the same conditions in our code and obtain $33$ synthetic lensed Quasars. In Fig. \ref{SDSS}, we show the histogram of the redshift distribution for our 33 synthetic lensed Quasars (dashed line) and give, for comparison, the redshift distribution for the 26 lensed Quasars  observed by the SDSS (continuous line). Overall, the two distributions appear in agreement even if there appear to be a small excess, in the simulated sample, at low reshifts.}

\begin{figure}[]
\includegraphics[width=10.5 cm]{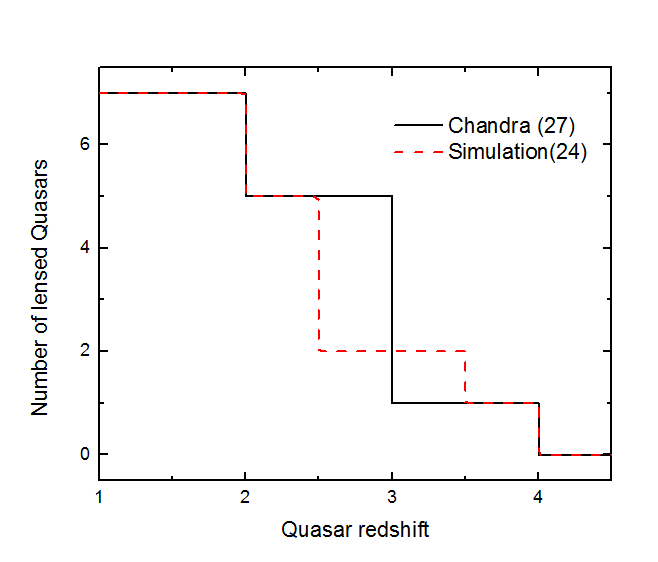}
\caption{Histogram of the redshift distributions for 24 synthetic lensed Quasars (dashed line) and  27 lensed Quasars observed by Chandra (continuous line).}
\label{Chandra}
\end{figure}

{In the case of Chandra observations we consider the results obtained by Walton et al. \cite{walton2015broad}, who analysed $27$ lensed Quasars observed by Chandra with redshift in the range  $1.0\leq z \leq 4.5$. In order to estimate the overall number of Quasars in the Chandra field of view, we consider the fact that it observed $320 ~deg ^2 $  of the sky up to a flux limit of $10^{-13}erg/s/cm^2$, $135~ deg^2$ up to a flux limit of $10^{-14} erg/s/cm^2$ and $6 ~deg^2$  up to a flux  limit of $10^{-15}erg/s/cm^2$ (see Evans et al. \cite{evans2010chandra}). Using the results obtain in \cite{hasinger2005luminosity} relative to the Quasar cumulative number counts N(>S) as a function of the flux limit, we estimate the total number of Quasars observed  by the Chandra telescope to be $12,725$. The Chandra angular resolution is about $0.5 ~$arcsec. \\
We then simulate the same conditions inside our code and find $24$ synthetic lensed Quasars with image separation beyond about than $0.5 ~$arcsec. In Fig. \ref{Chandra} we give the obtained histogram of the redshift distributions for 27 Lensed Quasars observed by Chandra  and 24 synthetic lensed Quasars simulated numerically. One can remark a good match in all bins of the Quasar redshift.}

\subsection{Prediction on IRT-THESEUS}

\begin{figure}[]
\includegraphics[width=10.5 cm]{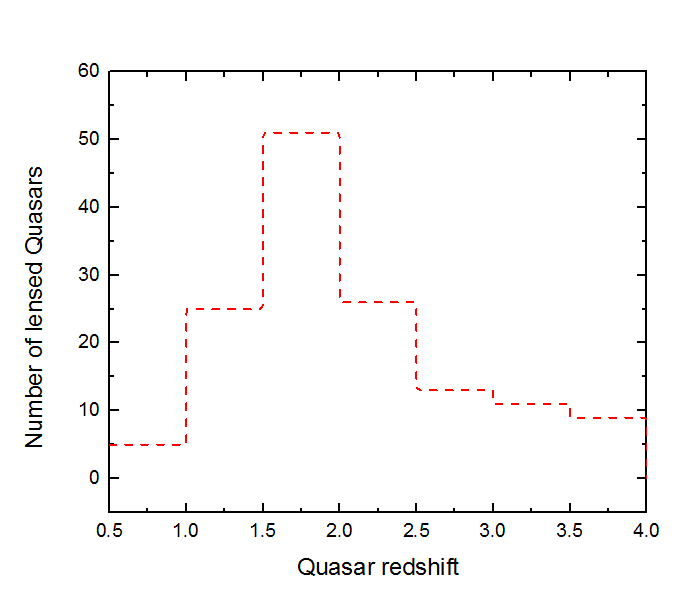}
\caption{Histogram of the redshift distribution for the 140 expected Quasars, lensed by foreground galaxies, and detectable by IRT-THESEUS.}
\label{Theseus}
\end{figure}

{Based on the number of Quasars ($\simeq 2.14\times 10^5$) expected to be detectable by IRT-THESEUS (see section 2) and requiring the angular separation  between images to be above the IRT-THESEUS angular resolution ($\simeq 1$ arcsec) \cite{amati2018THESEUS}, our algorithm produces about  $140$ strongly lensed Quasars by foreground galaxies.\\
In  Fig. \ref{Theseus}  the obtained  redshift distribution of the lensed Quasars is shown. As one can see, the redshift distribution peaks for redshift values in the range $1.5-2$ and becomes almost flat for Quasar redshifts larger than about 2.5.}

\section{Concluding remarks}

{THESEUS is an ESA space based project. Its specific scientific objective is to create a complete census of the GRB population in the first billion years since the Big Bang. During the time of waiting for any GRB alert, its three instruments will be able to create a rich database on objects at high $z$.\\
One of them is IRT-THESEUS, a 0.7 m class telescope with $15\times 15$ arcmin FOV, observing in the $0.7-1.8$ $\mu m$ band. 

In this work, we estimated the number of Quasars that IRT-THESEUS will be able to detect during the waiting time between GRB alerts, and found that about $2.14\times 10^5$ Quasars should be detected by the IRT instrument. Our calculation is based on the Quasar luminosity function (QLF) obtained by Lacy et al. \cite{lacy2015spitzer} from the Spizer Space Telescope imaging survey. A parallel calculation for Spitzer gives a lower number, $\simeq 2.55\times 10^4$, which is, in fact, close to the number $20,000$ of AGN candidates existing in the Spitzer archive \cite{lacy2020active}. \\
We have to remark here that we used the same QLF for both instruments, even if there is some discrepancy in their band observation, $0.7-1.8$ $\mu m$ for IRT-THESEUS and $3.6$ $\mu m$ for Spitzer, whereas the QLF is established for the waveband of $5$ $\mu m$.

The collection of data by IRT-THESEUS could help  in recognizing and studying in more details the Quasar population. An important phenomenon related to Quasars is their gravitational lensing by foreground galaxies.\\ Gravitational lensing is nowadays a powerful method of investigation both in astrophysics and cosmology for the study of Quasars, the distribution of stars in galaxies, in evaluating the amount  of the dark matter component in galaxies and even in discovering extragalactic planets.\\
Lensed Quasars are so far detected by several facilities, like Hubble Space Telescope (HST), SDSS, Chandra, Spitzer etc. Their observation in different bands give a  great help in understanding the physical processes that take place in Quasars.

Inspired by the great number of Quasars predicted to be observed by IRT-THESEUS we considered the possibility of observation of strongly lensed Quasars. For this reason, we establish Monte Carlo numerical simulations in order to estimate the expected number of   lensed Quasars detectable by IRT-THESEUS.\\
In our calculations we used the redshift distributions of both Quasars and galaxies, as well as the galactic mass distribution and stellar velocity dispersion (see  section 3).\\
First of all we compare the results of our numerical code on two existing databases: SDSS and Chandra X-Ray Telescope and find a close compatibility. Secondly, based on the estimated number of Quasars detectable by IRT-THESEUS ($\simeq 2.14\times 10^5$) and requiring the angular separation  between the lensed images to be above the IRT-THESEUS angular resolution ($\simeq 1$ arcsec), we find that about $140$ strongly lensed Quasars by foreground galaxies would be detectable by IRT-THESEUS.\\
Another instrument we were interested in is eROSITA European X-Ray instrument, which observes in ($0.5-2.0 ~$ keV band), with an angular accuracy of $15 ~$arcsec \cite{merloni2012erosita}, designed to observe about 3 million AGN in the extragalactic sky (|b| > 10°) \cite{kolodzig2013agn}. We checked  this number by using a similar calculation as in the case of Chandra, and found a very close value, $2.9\times 10^6$. Although this number is high, our algorithm does not produce any single lensed AGN by foreground galaxies in the case of eROSITA, due to relatively large angular accuracy.\\ 
We would like to remark that finding strongly lensed Quasars is of  great importance in astrophysics and cosmology since it allows increasing our knowledge not only of Quasars, but also of the structure,  dark matter content, dynamics and evolution, of the lens galaxies. It may even allow  further   test the theory of general relativity, constrain the parameters of  alternative gravity theories and gather information on the cosmological parameters.}

{\bf Acknowledgements}   
The authors are grateful to the anonymous  referee for his/her comments which significantly help to improve the manuscript. FDP and AAN acknowledge the TAsP and Euclid INFN Projects. 

\bibliography{main}

\end{document}